\documentclass[%
    reprint,
    aps,
    prl,
    superscriptaddress,
    showpacs,
    floatfix,
    bibnotes,
    amsmath,
    amssymb,
    amsfonts,
    amsthm,
]{revtex4-2}

\usepackage[utf8]{inputenc}     % Input encoding
\usepackage[T1]{fontenc}        % Font encoding
\usepackage{color}              % Colors
\usepackage[svgnames]{xcolor}   % More colors
\usepackage[inkscapelatex=true]{svg}
\usepackage{dcolumn}            % Align table columns on decimal point
\usepackage{bm}                 % Bold math
\usepackage{enumerate}          % Better enumeration
\usepackage{enumitem}	        % Styled enumerations
\usepackage{csquotes}           % Better quoting
\usepackage{empheq}             % Emphasize equations
\usepackage{dsfont}             % Used for double line symbols like R
\usepackage{siunitx}            % SI units
\usepackage{physics}            % Bra-Ket notation
\usepackage{hyperref}           % Add hypertext capabilities
\usepackage[caption=false]{subfig}
\usepackage{graphicx}           % Figures
\usepackage{cleveref}

\usepackage{todonotes}

\graphicspath{{figs/}}

\hypersetup{
	colorlinks  = true,
	citecolor   = Orange!80!black,
	linkcolor   = RoyalBlue!60!black, % DarkOrchid!70!black,
	urlcolor    = RoyalBlue!80!black
}
\usepackage{tikz}
\usetikzlibrary{arrows}

\tikzstyle{site} = [circle, shading=ball, ball color=black!40, minimum size=4mm, inner sep=0pt]

\begin{document}

%\preprint{APS/123-QED}
\title{Snapshot renormalization group for quantum matter}

\author{Laurin \surname{Brunner}}
\email{laurin.brunner@uni-a.de}
\affiliation{Theoretical Physics III, Center for Electronic Correlations and Magnetism, Institute of Physics, University of Augsburg, D-86135 Augsburg, Germany}

\author{Tobias \surname{Wiener}}
\email{tobias.wiener@uni-a.de}
\affiliation{Theoretical Physics III, Center for Electronic Correlations and Magnetism, Institute of Physics, University of Augsburg, D-86135 Augsburg, Germany}

\author{Tiago \surname{Mendes-Santos}}
\affiliation{Pasqal, Paris, France}

\author{Reyhaneh \surname{Khasseh}}
\email{reyhaneh.khasseh@uni-a.de}
\affiliation{Theoretical Physics III, Center for Electronic Correlations and Magnetism, Institute of Physics, University of Augsburg, D-86135 Augsburg, Germany}

\author{Markus \surname{Heyl}}
\email{markus.heyl@uni-a.de}
\affiliation{Theoretical Physics III, Center for Electronic Correlations and Magnetism, Institute of Physics, University of Augsburg, D-86135 Augsburg, Germany}
\affiliation{Centre for Advanced Analytics and Predictive Sciences (CAAPS), University of Augsburg, Universitätsstr. 12a, 86159 Augsburg, Germany}

\date{\today}
%\doi{...}
%\pacs{...}
 % Use showkeys class option if keyword display desired

\begin{abstract}
Recent advances in quantum simulator experiments enable unprecedented access to quantum many-body states through snapshot measurements of individual many-body configurations.
Here, we introduce an \textit{exact} renormalization group (RG) transformation that can be directly applied to any such snapshot dataset.
Our SnapshotRG operates in real space, but can also be directly translated to an RG in the abstract dataspace of measurement configurations, providing a framework for the characterization of quantum many-body systems on a more general level.
We demonstrate that snapshot datasets in dataspace exhibit self-similarity at continuous phase transitions, providing an explanation for the recently observed scale-freeness of so-called wavefunction networks.
As a consequence, scale invariance extends beyond traditional low-order correlation functions to encompass the full statistical structure of quantum states as contained in their snapshot datasets.
Our SnapshotRG can be readily implemented with snapshot data generated by numerical method such as neural quantum states or any quantum simulation platform, offering a versatile tool for characterizing quantum phase transitions and critical phenomena in quantum matter.
\end{abstract}

\maketitle

%\tableofcontents

% =====================================================================================================================
% INTRODUCTION
% =====================================================================================================================

\paragraph{Introduction.}
\label{sec:introduction}

The impressive progress in experimental quantum systems over the past two decades has lead to unprecedented control over quantum many-body systems~\cite{Gross2017, Browaeys2020, Monroe2021}.
A key development in modern quantum simulator and quantum computing platforms is the capability to perform projective snapshot measurements of many-body quantum states~\cite{Sherson2010, Bakr2009, Gross2021, Asteria2021, Haller2015}.
These snapshots correspond to joint projective measurements on each quantum degree of freedom yielding as the measurement outcome a full many-body configuration.
While this experimental progress has created new opportunities for probing quantum matter, it simultaneously presents the challenge of extracting physical insights from the resulting high-dimensional datasets without an a-priori information loss, e.g., through a dimensional reduction to low-order correlation functions~\cite{Brydges2019, Bohrdt2019, Bohrdt2021, Elben2020, Vanoni2024}.
Traditional approaches based on full quantum state tomography become exponentially impractical for large systems, necessitating the development of data analysis methods that can efficiently extract relevant information about quantum phases, phase transitions, and critical phenomena directly from measurement snapshots.
Among emerging approaches, wave function networks (WFN) offer a promising framework by treating the measurement dataset as a graph and analyzing its structural properties to extract physical insights~\cite{mendes-santos_wave-function_2024, iakovlev_WFN}.
Building on such graph-based and other data-driven methodologies, there is a broader need to develop comprehensive theoretical frameworks that can systematically bridge the gap between raw experimental data and fundamental physical understanding, enabling the characterization of quantum matter without requiring complete knowledge of the underlying quantum state.

\begin{figure*}
    \centering
    \includegraphics[width=0.98\linewidth]{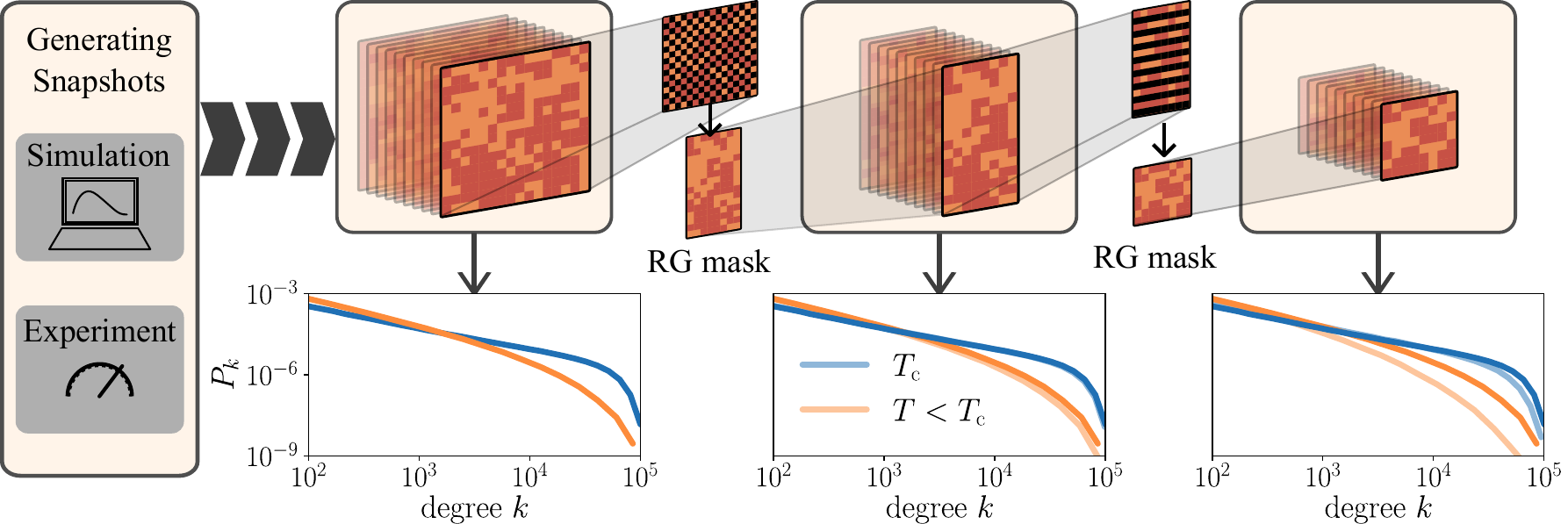}
    \caption{Workflow of the SnapshotRG method on a 2D simple cubic lattice, illustrated for two consecutive RG steps.
    Snapshots are obtained from either simulations or experiments and processed through iterative SnapshotRG steps.
    Each SnapshotRG step acts as a mask on individual snapshots, effectively removing every second site.
    After the first RG transformation, the system becomes a simple cubic lattice with rescaled lattice constant rotated by 45°, requiring every second RG step to be performed in a rotated coordinate system.
    The degree distribution of the WFN, calculated at each SnapshotRG step independelty, remains invariant at phase transitions but changes away from critical points.}
    \label{fig:rg_transformation}
\end{figure*}

In this work we introduce an exact renormalization group (RG) transformation formulated directly on snapshot datasets.
This SnapshotRG eliminates in one RG step every second lattice site in real space (see Fig.~\ref{fig:rg_transformation}), following conventional real-space decimation RG methods~\cite{kadanoff_rg}.
What is different, however, is that we don't determine the renormalized real-space Hamiltonian after performing an RG step, which generally requires some perturbative approximation schemes.
Instead, we directly determine the impact of integrating out every second lattice site on the level of the snapshot datasets taken on the reduced set of degrees of freedom.
This amounts to studying the impact of a real-space RG on the Born probability from which the snapshots are obtained as samples.
We show that this SnapshotRG can be determined from a snapshot dataset taken for the original system including all lattice sites by masking out for each snapshot those degrees of freedom which are integrated out, see Fig.~\ref{fig:rg_transformation}.
Consequently, this proposed RG can be executed directly on any existing snapshot dataset taken in theory or experiment.
Importantly, we demonstrate that the SnapshotRG cannot only be associated with an exact conventional real-space RG, but also as an RG operating in the dataspace of snaphots.
This is of particular importance in view of the recently uncovered scale-freeness~\cite{mendes-santos_wave-function_2024, Albert2002, barabasi_scalefree} of WFNs at continuous phase transitions: we find that the underlying degree distribution of WFNs remains invariant under the RG.
Overall, the introduced SnaphotRG is a universally applicable tool for snapshot datasets taken in experiments and theory, enabling a general RG analysis not only on conventional observables but on a more general level for the quantum state.

\paragraph{SnapshotRG.}

Exact real-space renormalization group (RG) transformations for interacting spin models beyond one dimension often present formidable theoretical and computational challenges that render them impractical for direct implementation.
However, we show that, when working on the level of snapshot measurements, such exact RG transformations become computationally tractable and conceptually straightforward.
Following the real-space renormalization group approach originally proposed by~\cite{kadanoff_rg}, we implement an RG transformation by systematically integrating out every second degree of freedom on a lattice, see Fig.~\ref{fig:rg_transformation}.

To formalize the approach, consider a quantum spin-1/2 system defined on subsystems A and B, characterized by a density matrix $\rho$.
A snapshot $|\mathbf{s}_\mathrm{A}\mathbf{s}_\mathrm{B}\rangle$ in the $\sigma^z$-basis is drawn with probability $P_\mathrm{phys}(|\mathbf{s}_\mathrm{A}\mathbf{s}_\mathrm{B}\rangle) = \mathrm{Tr}\left(\rho |\mathbf{s}_\mathrm{A}\mathbf{s}_\mathrm{B}\rangle\langle \mathbf{s}_\mathrm{A}\mathbf{s}_\mathrm{B}|\right)$ yielding a snapshot data set $\left\{|\mathbf{s}_\mathrm{A}\mathbf{s}_\mathrm{B}\rangle \right\}_{P_\mathrm{phys}}$.
The SnapshotRG procedure discards the degrees of freedom belonging to B, mapping every full snapshot $|\mathbf{s}_\mathrm{A}\mathbf{s}_\mathrm{B}\rangle$ to its reduced form $|\mathbf{s}_\mathrm{A}\rangle$, thereby generating a coarse-grained dataset $\left\{|\mathbf{s}_\mathrm{A}\rangle \right\}_{P'}$.
The probability of obtaining a particular reduced snapshot $|\mathbf{s}_\mathrm{A}\rangle$ after the SnapshotRG is given by

\begin{equation}
    P'(\mathbf{s}_\mathrm{A}) = \sum_{\mathbf{s}_\mathrm{B}} P_\mathrm{phys}(\mathbf{s}_\mathrm{A}\mathbf{s}_\mathrm{B}) = \mathrm{Tr}(\rho_\mathrm{A} |\mathbf{s}_\mathrm{A}\rangle\langle \mathbf{s}_\mathrm{A}|) = P_\mathrm{phys}(\mathbf{s}_\mathrm{A})
\end{equation}
where $\rho_\mathrm{A} = \mathrm{Tr}_\mathrm{B}(\rho)$ is the reduced density matrix.
This identity holds because the local computational basis satisfies the decomposition $\sum_{\mathbf{s}_\mathrm{B}} |\mathbf{s}_\mathrm{A}\mathbf{s}_\mathrm{B}\rangle\langle \mathbf{s}_\mathrm{A}\mathbf{s}_\mathrm{B}| = |\mathbf{s}_\mathrm{A}\rangle\langle \mathbf{s}_\mathrm{A}| \otimes\sum_{\mathbf{s}_\mathrm{B}} |\mathbf{s}_\mathrm{B}\rangle\langle \mathbf{s}_\mathrm{B}|$.
Consequently, the post-RG snapshot dataset follows precisely the same probability distribution that would be obtained by first computing the reduced density matrix and then sampling snapshots from it.
The mathematical equivalence demonstrates that our decimation procedure in measurement space corresponds exactly to the partial trace operation in the quantum mechanical description.
Consequently, the coarse-grained snapshots obtained through spin removal represent faithful samples from the renormalized quantum state, preserving all statistical properties relevant for the RG flow.
In practice, the SnapshotRG transformation can be implemented straightforwardly by applying a binary mask to each snapshot configuration, effectively selecting which spin degrees of freedom to retain in the coarse-grained representation.
This transformation operates directly on the measurement data rather than on the theoretical Hamiltonian, thus circumventing the usual approximations required in most conventional RG treatments.

\paragraph{Model.}
To highlight the capabilities of our SnapshotRG, we generate snapshot data sets with numerical simulations of the well-known transverse field Ising model (TFIM) in two dimensions (2D) and three dimensions (3D), which is described by the Hamiltonian $\hat{H} = -J \sum_{\langle i,j\rangle} \sigma_i^z \sigma_j^z + g \sum_i \sigma_i^x$, where $J$ is the spin coupling, $g$ is the strength of the transverse field and $\langle .,.\rangle$ denotes nearest neighbor terms.
In particular, we will consider the system in the vicinity of continuous thermal and quantum phase transitions in different dimensions to identify universal structures in snapshot data sets.
The TFIM undergoes a quantum phase transition at a critical $g_\mathrm{c}/J\approx 3.044$ (for 2D)~\cite{Rieger1999, Mondaini2016}.
Configuration snapshots are obtained by sampling from the groundstate according to the Born rule $P(\mathbf{s})=|\psi_0(\mathbf{s})|^2$ using Neural Quantum States (NQS)~\cite{carleo_NQS, schmitt_NQS}, which provide an efficient variational representation of the many-body wavefunction and will be described in more detail in the End Matter.

Since thermal phase transitions are fundamentally classical phenomena driven by temperature-induced fluctuations~\cite{Sachdev2011}, we generate snapshot datasets by Monte Carlo sampling configurations from classical Hamiltonian systems at finite temperature.
For the classical thermal transitions, we consider large scale (up to $N = 2^{16}$ sites) Ising models $H_0 = -J \sum_{\langle i,j\rangle} s_i s_j$ in both two and three spatial dimensions.
Configuration snapshots are obtained through Monte Carlo simulations that implement a Markov chain sampling process to draw configurations from the canonical thermal equilibrium distribution
\begin{equation}
P(\mathbf{s}) = Z^{-1} \exp(-\beta H(\mathbf{s}))
\end{equation}
where $Z$ is the partition function and $\beta = 1/(k_B T)$ is the inverse temperature.
We sampled snapshots at the respective critical temperature $T_\mathrm{c}$ ($=\frac{2J}{k_\mathrm{B}\ln(1+\sqrt{2})}$ for 2D~\cite{Onsager1944} and $\approx4.512 J/k_\mathrm{B}$ for 3D~\cite{Hasdenbusch2010, Blote1995}) where the systems undergo second-order phase transitions.
To mitigate the critical slowing down phenomenon that occurs near phase transition points, we combine standard Metropolis single-spin updates with cluster-based Wolff algorithm~\cite{Wolff_Updates} moves.

\begin{figure}
    \centering
    \includegraphics[width=1\linewidth]{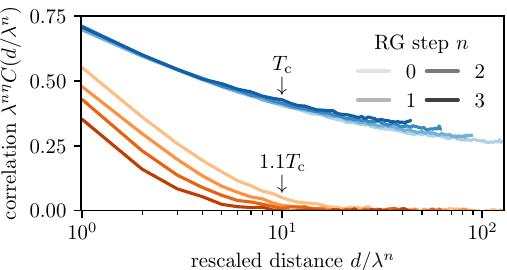}
    \caption{
    Effect of SnapshotRG transformation on rescaled spin-spin correlation $C(d) = \frac{1}{L^2}\sum_i\langle s_i s_{i+d}\rangle$ of the classical 2D Ising model.
    $\lambda=\sqrt{2}$ is the rescaling factor and $\eta=\frac{1}{4}$ is the critical exponent that describes the power-law decay of the correlation function at the critical point.
    At the critical temperature $T_c$ the SnapshotRG leaves the correlation invariant while for higher temperatures there is a RG flow to infinite temperature.
    }
    \label{fig:RG_correlations}
\end{figure}

Let us first take the chance to demonstrate the working principle of the SnapshotRG for conventional correlation functions such as $C(d) = \frac{1}{L^2}\sum_i\langle s_i s_{i+d}\rangle$.
Concretely, we show in Fig.\ref{fig:RG_correlations} the dependence of $C(d)$ upon applying the RG transformation for the 2D Ising model.
Here, we have rescaled both real-space distances and the correlation function by the rescaling factor $\lambda=\sqrt{2}$, which measures the change of lattice spacing by performing one RG decimation step on the 2D square lattice.
At the critical temperature $T=T_c$ the rescaled correlation function exhibits a data collapse as expected from scale invariance and universality.
For the exemplary temperature $T=1.1T_c$ on the other hand, the correlation function transforms to lower and lower correlation length under the RG consistent with flowing to the infinite-temperature fixed point expected for $T>T_c$.

\paragraph{Wave function networks.}
After having introduced the basic idea of the SnapshotRG procedure and after demonstrating its performance on conventional correlation functions, we now take the next step by showing that it can also be utilized as an RG in the dataspace of snapshot measurements.
Concretely, we will consider the recently introduced wave function networks (WFN) generated from snapshot datasets, which have been shown to exhibit universal scale-free features~\cite{mendes-santos_wave-function_2024}.

A WFN treats each individual snapshot as a node in an undirected graph embedded in a high-dimensional space and connects two nodes if their Hamming distance $D(\mathbf{s}, \mathbf{s}') = \frac{1}{2} \sum_i |s_i - s'_i|$ is below a suitably chosen cutoff $R$.
A more detailed description of the construction of WFNs can be found in the end matter.
A fundamental characteristic in network theory is the degree distribution, which describes the statistical distribution of node connectivities throughout the network.
The degree of a node is defined as the total number of edges connecting it to other nodes in the graph.
Scale-free networks~\cite{barabasi_scalefree, barabase_scalefree2} represent a particularly important class of complex networks where the degree distribution follows a power-law relationship of the form 
\begin{equation}
    P_k \sim k^{-\gamma}, 
\end{equation}
where $k$ is the degree and $\gamma$ is the scaling exponent.
This power-law behavior indicates the presence of highly connected hub nodes alongside a majority of sparsely connected nodes, reflecting underlying hierarchical or critical organization.
In contrast, networks with random connectivity typically exhibit degree distributions that approximate a Poisson distribution.
The distinction between power-law and Poisson degree distributions serves as a diagnostic tool for identifying the presence of emergent structure and correlations within the underlying system generating the network.

\paragraph{Universality in snapshot datasets.}
\label{sec:results}

\begin{figure}
    \centering
    \includegraphics[width=0.98\linewidth]{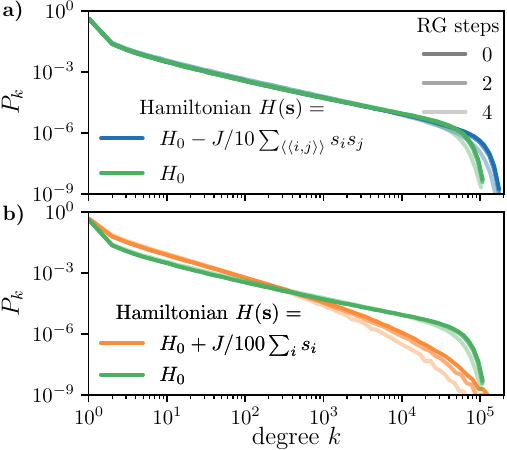}
    \caption{SnapshotRG transformation on the degree distribution $P_k$ for the 2D Ising model in the presence of perturbations.
    Initial system size before the SnapshotRG is $L_x\times L_y = 256\times 256$.
    Number of samples for all curves is $N_r = 5\cdot10^5$.
    \textbf{a)} Symmetry-preserving perturbation. The degree distribution remains invariant under the SnapshotRG.
    For the perturbed system the temperature is chosen such that the system is located at the phase transition point.
    \textbf{b)} Symmetry-breaking perturbation. The degree distribution of the perturbed system exhibits a different shape without clear power-law and ceises to be an invariant under the SnapshotRG.
    The perturbed system is chosen at the critical temperature of the unperturbed system.}
    \label{fig:hamiltonian_pertubations}
\end{figure}

One key numerical observation for the studied Ising systems is that the degree distribution $P_k$ not only shows scale-free behavior (power-law decay) at continuous phase transitions but is also invariant under the SnapshotRG transformation (see, e.g., ~\ref{fig:rg_transformation}\&\ref{fig:hamiltonian_pertubations}).
Consequently, these systems exhibit a form of scale invariance on the level of the entire snapshot dataset.
To test this consideration we study further the behavior of the degree distribution $P_k$ in the vicinity of continuous phase transitions in the Ising model in the presence of two qualitatively distinct perturbations, see Fig.~\ref{fig:hamiltonian_pertubations}.
The first perturbation $V_1 = - J/10 \sum_{\langle \langle i,j\rangle\rangle} s_i s_j$  incorporates additional next-nearest neighbor interactions while preserving all fundamental symmetries of the original Ising model, ensuring that both systems belong to the same universality class.
The second perturbed model $V_2 = J/100 \sum_i s_i$ introduces a longitudinal magnetic field term that explicitly breaks the $\mathbb{Z}_2$ symmetry of the Ising model, thereby introducing a relevant perturbation.
According to the universality hypothesis, physical systems within the same universality class exhibit identical critical exponents and scaling behaviors near phase transitions, regardless of microscopic details.
Our analysis reveals that the degree distributions obtained from wavefunction networks of $H_0$ and $H_0 + V_1$ remain essentially unchanged, confirming that the network properties are insensitive to weak symmetry-preserving perturbations, which are considered irrelevant in the RG sense.
In stark contrast, the degree distribution for model $H_0 + V_2$ exhibits qualitatively different behavior, reflecting the departure from universality induced by the symmetry-breaking magnetic field, which constitutes a relevant perturbation.
This comparison establishes that wavefunction network degree distributions serve as effective probes of universal critical phenomena also on the level of the snapshot dataspace, distinguishing between systems based on their fundamental symmetries rather than specific microscopic interactions.
Moreover it is universal information contained not only in expectation values of local observables, but rather of the entire data structure of snapshots codified by the WFN.

\paragraph{Critical exponents.}

\begin{figure}[htbp]
    \centering
    \includegraphics[width=1\columnwidth]{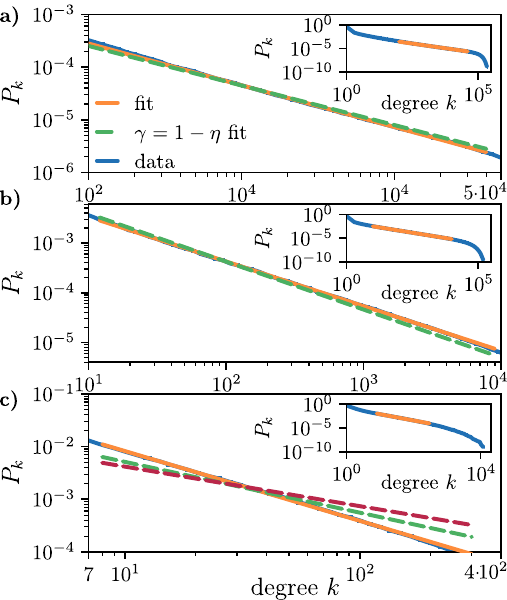}
    \caption{The power-law of the degree distribution at critical points for \textbf{a)} 2D classical Ising model (square lattice with $L = 256$), \textbf{b)} 3D classical Ising model (cubic lattice with $L= 40$), and \textbf{c)} 2D quantum transverse field Ising model (square lattice with $L = 16$).
    For all lines we took $N_r=10^6$ snapshots.
    The extracted exponents are \textbf{a)} $\gamma = -0.79 \pm 0.02$, \textbf{b)} $\gamma = -0.90 \pm 0.01$, and \textbf{c)} $\gamma = -1.32 \pm 0.01$.
    Reference power-law curves with exponent $\gamma = 1 - \eta$ are included for comparison, where $\eta$ represents the corresponding critical exponent of each system.
    For panel \textbf{c)}, additional reference curves using the $\eta$ values from both two- and three-dimensional classical Ising models are displayed, demonstrating that neither provides agreement with the quantum system behavior.}
    \label{fig:critical_exponents}
\end{figure}

The universal behavior and self-similarity observed in the degree distribution of WFNs under RG transformations suggest a fundamental connection to the critical exponents that characterize phase transitions.
Since prior research on WFNs did not settle the power-law exponents $\gamma$ of the degree distribution $P(k) \sim k^{-\gamma}$, we investigate them at the critical point for classical Ising models in two and three dimensions, as well as for the 2D quantum transverse-field Ising model.
For an intermediate range of $k$ values a power-law scaling is clearly observed, which enables us to fit over multiple orders of magnitude ensuring robust extraction of the scaling exponent $\gamma$.  
The results of this analysis are presented in Figure \ref{fig:critical_exponents}, which demonstrates the systematic variation of the degree distribution exponent across different system dimensionalities and transition types.
For thermal phase transitions in both two and three dimensions, we observe a compelling connection between the network exponent $\gamma\approx1-\eta$ and the critical exponent $\eta$.
However, the quantum phase transition in the 2D transverse field Ising model does not exhibit this same relationship.
This deviation from the classical scaling behavior may be attributed to finite-size effects that become particularly pronounced in the quantum system, where the accessible system sizes are typically smaller than their classical counterparts due to computational constraints.

\paragraph{Conclusion}
\label{sec:conclusion}
In this work we have introduced the SnapshotRG framework, that operates directly in the space of many-body measurement snapshots and provides an exact real-space decimation procedure without requiring an analytical reformulation of the Hamiltonian. 
This data-driven approach reveals scale invariance not only in traditional observables such as correlation functions, but extends to the entire dataspace through WFNs, whose degree distributions $P_k$ remain invariant under the SnapshotRG at criticality. We find numerical evidence for the considered thermal phase transitions that the power-law exponent $\gamma \approx 1 - \eta$ of $P_k$ is related to the critical exponent $\eta$ of the underlying transition, while quantum critical points appear to behave differently, which remains an open question for the future.

The SnapshotRG is a powerful, ready-to-use tool that allows existing and future quantum simulator snapshot data to be analyzed for RG flow and universal behavior directly, opening a new avenue for data-driven discovery in experimental many-body physics.
On an equal level, the SnapshotRG can be applied to theoretical snapshot data as readily available in neural quantum state simulations for instance, as we have demonstrated in the present letter.
For the future it would be an interesting open question to which extent the introduced SnapshotRG can also reveal insights on other methods for quantifying snapshot datasets, such as the intrinsic dimension or Kolmogorov complexity~\cite{Mendes-Santos_2021}.
%

% =====================================================================================================================
% APPENDIX
% =====================================================================================================================

\begin{acknowledgments}
\paragraph{Data availability - } The data contained in all figures of this article is available on Zenodo~\cite{Zenodo}.

\paragraph{Acknowledgements - }
We thank Andreas Elben for insightful discussions.
The authors gratefully acknowledge the resources on the LiCCA HPC cluster of the University of Augsburg, co-funded by the Deutsche Forschungsgemeinschaft (DFG, German Research Foundation) – Project-ID 499211671.
This project has received funding from the European Research Council (ERC) under the European Union’s Horizon 2020 research and innovation programme (grant agreement No. 853443), and by the research funding program "Forschungspotenziale besser nutzen!" of the University of Augsburg.
This work was in part supported by the Deutsche Forschungsgemeinschaft under grant FOR 5522 (project-id 499180199).

\end{acknowledgments}

% ~\nocite{*}
\bibliography{sources} % Produces the bibliography via BibTeX.

\newpage
\appendix
\section{End Matter}

\begin{table}[h]
\centering
\begin{tabular}{ll}
\hline
\textbf{Markov Chain (Optimization)}           & \\
\hline
Number of Samples               & 5000 \\
Sweep Steps                     & 256 \\
Thermalization Sweeps          & 1 \\
Number of Chains    & 500 \\
\hline
\textbf{Markov Chain (Snapshot Generation)}           & \\
\hline
Number of Snapshots               & $10^6$\\
Sweep Steps                     & 1024 \\
Thermalization Sweeps          & 5 \\
Number of Chains    & 500 \\
\hline
\textbf{TDVP}                   & \\
\hline
Initial Diagonal Shift          & 10 \\
Shift Interval                  & 200 \\
Shift Factor                    & $10^{-1}$ \\
Minimum Diagonal Shift          & $10^{-10}$ \\
Convergence Threshold           & $10^{-6}$ \\
Euler Time Step                 & $10^{-2}$ \\
\hline
\textbf{ResNet}                 & \\
\hline
Filter Size                     & $3$ \\
Channels                        & 8 \\
Number of Blocks                        & 4 \\
Strides                         & 1 \\
Bias                            & True \\
\hline
\end{tabular}
\caption{Hyperparameters for the NQS method.}
\label{tab:hyperparameters}
\end{table}

We employ the neural quantum state (NQS)~\cite{carleo_NQS, schmitt_NQS} method to investigate the quantum transverse-field Ising model (TFIM).
The NQS approach provides an approximation to the ground state wavefunction as $|\psi_0\rangle \approx \sum_{{\mathbf{s}}} \psi_{\theta} (\mathbf{s})|\mathbf{s}\rangle$, where $\psi_\theta(\mathbf{s})$ is a neural network parametrized by variational parameters $\theta$.

The specific neural network is a convolutional neural network (CNN) architecture based on the ResNet framework~\cite{he_ResNet, chen_ResNet} with Gaussian Error Linear Unit (GELU) activation functions~\cite{hendrycks2023}.
The network incorporates residual skip connections to enhance numerical stability during training and uses exclusively real-valued parameters.
The architecture consists of $n$ sequential blocks, where each block contains two convolutional layers with GELU activation functions applied between them.
The convolutional kernels have spatial dimensions $F \times F$ with filter depth $C$.
The specific hyperparameters used are listed in Table~\ref{tab:hyperparameters}.
Following the final block, the real-valued output tensor is partitioned into two equal parts.
The first partition represents the real component of the complex-valued output, while the second partition represents the imaginary component.
This splitting mechanism enables the network to produce complex-valued predictions while using well-established activation functions for real valued inputs.

The variational network parameters are optimized using stochastic reconfiguration.
For numerical stability, we employ a diagonal shift in the Hamiltonian, reducing it over the course of training until reaching a minimum value.
The TDVP equations are integrated using an Euler scheme.
A Markov chain Monte Carlo (MCMC) sampling scheme is employed to generate samples from the $|\psi(s)|^2$ distribution, required for both the TDVP equation and snapshot generation.
The MCMC hyperparameters are chosen differently for optimization and snapshot generation phases.
The snapshots used in the main text are generated after the optimization converged quantified by an energy variance density below $10^{-6}$.

We used the jVMC Codebase to carry out our numerical simulations ~\cite{schmitt_jvmc}.
All relevant hyperparameters can be found in Table~\ref{tab:hyperparameters}.

\paragraph{Wave function network construction}

This discussion follows \cite{mendes-santos_wave-function_2024}.
We consider the case of spin-1/2 systems where each snapshot can be represented as a configuration of $N$ spin variables $\{s_l\}_{l\in [1, N]}$, where $s_l \in \{-1, +1\}$ denotes the state of the $l$-th spin.
Furthermore we require that each snapshot in the dataset is unique, since duplicates can skew the calculation of the following cutoff $R$ and degrees $k$.
From the set of $N_r$ unique snapshots, we construct an undirected graph in the $2^N$-dimensional configuration space, treating each snapshot as a node.
Two nodes are connected if their Hamming distance $D(\mathbf{s}, \mathbf{s}') = \frac{1}{2} \sum_l |s_l - s'_l|$ is less than a cutoff threshold $R$.
We choose $R$ as the average nearest neighbor distance 
\begin{equation}
    R =N_r^{-1}\sum_{n=1}^{N_r} r_1^{(n)}~,
\end{equation}
where $r_1^{(n)}$ is the nearest-neighbor distance for snapshot $n$.
This choice of $R$ accounts adapts to the structure of the dataset.
The degree $k$ of each node is the number of its connections, and the degree distribution $P_k$ is estimated through histogram analysis.
To circumvent issues with finite sample sizes, which can distort the distribution's tail, we employ logarithmic binning, where degrees are grouped into bins with exponentially increasing size.

This network construction methodology is applied independently to each dataset obtained after successive RG transformations, allowing us to track how the network topology evolves under the RG flow.

\end{document}